\documentclass[%
reprint,
superscriptaddress,
 amsmath,amssymb,
 aps,
 prl,
showkeys
]{revtex4-2}

\usepackage{graphicx}
\usepackage{enumerate}
\usepackage{hyperref}
\usepackage{multirow}
\usepackage{comment}

\usepackage{tabularray}

\usepackage{xcolor}

\begin{document}

%\preprint{}

\title{A review on the use of complex networks in science education research}
% \author{Paula Tuz\'on$^1$, Antoni Salv\`a Salv\`a$^2$ and Juan Fern\'andez-Gracia$^3$}
% \author{Paula Tuz\'on}
% \email[Contact:]{juanf@ifisc.uib-csic.es}
% \email[]{Your e-mail address}
%\thanks{}
% \affiliation{Science Education Department, University of Valencia. Avda. Tarongers 4, 46022 Valencia, Spain}

\author{Paula Tuzón}
\email[Contact:]{paula.tuzon@uv.es}
\affiliation{Departamento de Didáctica de las Ciencias Experimentales, Universidad de Valencia, 46022, Valencia, Spain}

\author{Juan García-Castillo}
% \email[Contact:]{jantonio@ifisc.uib-csic.es}
\affiliation{Instituto de Física Interdisciplinar y Sistemas Complejos IFISC (CSIC-UIB), Campus UIB, 07122, Palma de Mallorca, Spain}

\author{Juan Fern\'andez-Gracia}
\email[Contact:]{juanf@ifisc.uib-csic.es}
\affiliation{Instituto de Física Interdisciplinar y Sistemas Complejos IFISC (CSIC-UIB), Campus UIB, 07122, Palma de Mallorca, Spain}

% \date{}

\begin{abstract}
Network-based approaches have become increasingly prominent in science education research as tools for analysing relational structures in learning, teaching, and knowledge production. This review presents a PRISMA-informed scoping analysis of 82 articles published in nine leading science education journals, which are organised into four main categories: concept networks, social networks, bibliographic networks, and attitudes or behavioural networks. We observe a sustained exponential growth in the use of network methods, indicating a still-emerging and expanding research area. Concept networks dominate the literature, followed by social network analyses linking interaction structure to learning outcomes and persistence, while bibliographic and abilities-oriented networks provide complementary meta-level and practice-focused perspectives. In addition, analysis of the coauthorship network reveals a highly fragmented field, characterised by many small and weakly connected research groups, typically organised within single application categories. Complementary analysis of a citation network that includes all referenced authors shows that, despite this limited collaboration structure, the field is intellectually organised around several major traditions—network science methodology, learning sciences, and argumentation in science education—linked by a small number of bridging authors. Overall, the literature remains largely descriptive, relying on static, single-layer representations and a narrow set of network metrics. We identify substantial opportunities for advancing science education research through stronger theoretical integration and the adoption of dynamic, multilayer, and coevolutionary network frameworks.
\end{abstract}

% \pacs{}

\maketitle


\section{Introduction}

Over the past two decades, the concept of “networks” has gained significant traction across disciplines, from physics and biology to sociology and education. In science education research, networks offer both a theoretical lens and a methodological tool to examine a wide range of phenomena—how students relate concepts, how teachers interact, how curricular components are connected, and how research fields themselves are structured. This review aims to critically examine how network theory and network-based methods have been used in science education, focusing on both empirical studies and theoretical contributions.

Our approach involves a systematic selection of papers published in nine of the most prominent journals in the field. We include only those studies where network theory or tools play a substantial analytical or conceptual role. Rather than limiting ourselves to one form of network (e.g., social networks or concept maps), we consider the broad range of applications: from qualitative concept networks and epistemic network analysis (ENA) to co-authorship and bibliometric mappings. By doing so, we seek to understand not only how networks are used, but also what kinds of educational questions they help answer, what assumptions they carry, and how their use has evolved.

To structure the review, we propose a categorization of the literature into four distinct types (see Tab.~\ref{tab:categories} and Fig.~\ref{fig:categories}), further subdivided into ten subcategories, based on what is being networked and for what purpose. These range from networks of student discourse and knowledge structures, to networks used to analyze curricula, assessments, or the structure of science education research itself. We find that while network methods are increasingly common, they are unevenly distributed across topics and often serve more as exploratory or descriptive tools than as vehicles for theory testing or hypothesis-driven inquiry.

This review contributes a synthetic overview of how network-based reasoning has shaped science education research, clarifies conceptual boundaries across different uses of “networks,” and suggests new directions for deeper integration of network science into educational theory and practice.

\section{Methodology}

\subsection{PRISMA}

This review follows a PRISMA-informed scoping methodology aimed at systematically mapping how network science approaches have been used within science education research. Scoping reviews are appropriate for fields that are conceptually diverse and methodologically heterogeneous, as is the case for network-based studies in science education, where the objective is to characterize the breadth, nature, and evolution of the literature rather than to conduct an effect-size synthesis. PRISMA principles were applied to ensure transparency in identification, screening, and inclusion procedures, while retaining the flexibility needed for comprehensive coverage.

The corpus was delimited to nine journals recognized as central publication venues in science education: PRPER (Physical Review Physics Education Research), Science Education, Science $\&$ Education, Research in Science Education, Journal of Research in Science Teaching, International Journal of Science Education, International Journal of Science and Mathematics Education, Journal of Science Education and Technology, and Instructional Science. These journals consistently occupy top positions (Q1–Q2) in major indexing systems such as Journal Citation Reports and Scimago, and collectively shape the mainstream discourse of science education research. Additionally we added the papers published in a special issue about complex networks in science education research~\cite{Koponen2020}.

The search was conducted between January and July 2025 and covered all publication years available in each journal up to that point. Instead of relying on predefined keyword strings, we used a two-stage, exhaustive strategy. First, we queried each journal’s internal search engine with the anchor term complex networks, which typically retrieves papers using network models even when the expression does not appear in titles or keywords. All records obtained in this step were examined in full. Second, to avoid missing studies whose methods might not be indexed under network-related terminology, we manually browsed every issue of every journal, screening titles, abstracts, and when necessary full texts. Articles were excluded only after full-text review confirmed that no graph-theoretical network model was actually used, or when terms such as network or neural network appeared only metaphorically or solely in reference to generic machine-learning classification. Neural network models were included only when explicitly used as structural or theoretical models of cognition or reasoning.

Inclusion criteria required that a study (1) be published in one of the nine selected journals; (2) employ a network model such as social, conceptual, semantic, epistemic, bibliometric, co-occurrence, or neural-connectivity networks; (3) address a topic directly relevant to science education; and (4) be published at any time up to July 2025. Exclusion criteria applied to studies using network terminology solely in metaphorical ways, to machine-learning papers using neural networks only as prediction engines, and to papers outside the scope of science education even if they employed network methods.

Screening proceeded in three stages consistent with PRISMA recommendations: identification through search engines and manual browsing; full-text screening; and eligibility assessment. A PRISMA-style flow diagram may be used to summarize the number of records identified, screened, excluded, and retained, although quantitative synthesis was not required due to the scoping nature of the review.

For each included article, information was extracted regarding the type of network model employed, the object of analysis (for example, students, teachers, curricula, reasoning processes), the methodological approach (such as social network analysis, epistemic network analysis, module analysis, co-word analysis, neural connectivity, or multimodal correlation networks), the educational domain, and the educational level. These dimensions informed the categorization scheme used in the review.

A scoping design was chosen because the use of complex networks in science education spans conceptual, social, epistemic, multimodal, and bibliometric approaches that differ substantially in purpose, methodology, and data sources. The aim of the review was therefore to document how and why network methods are being used across the field, to identify thematic and methodological clusters, and to synthesize emerging trends. PRISMA principles provided the transparency and procedural clarity required for a systematic review, while the scoping framework enabled a broad, integrative mapping of an evolving research landscape.

\subsection{Co-authorship and citation network}

To further characterise the structure and internal organisation of the field, we constructed a coauthorship network from the articles in our corpus. For each pair of authors, we computed the number of papers they co-authored, which we used as the weight of the corresponding (undirected) edge. To assess the overall cohesion of the field, we examined the network’s connectance and basic component structure: number of connected components, their sizes, and how these components relate to topical categories. Each author was assigned the set of categories associated with the papers they contributed to, and we quantified the diversity of categories present within each component.

To complement the coauthorship perspective, we constructed a citation network by adding all authors mentioned in the references of every paper in the corpus. For each ordered pair of authors, we counted how many times one cites the other within the dataset; however, for the analysis we disregarded directionality and treated the resulting network as undirected and weighted. This allowed a unified comparison with the coauthorship structure. We computed the same set of centrality measures on this citation network to highlight authors who are influential from a citation-flow perspective.

All network calculations were performed using the Python library NetworkX.

\section{Types of studies using complex networks in science education: summary}

Based on the search strategy described in the Methodology section and the systematic examination of the selected science education journals, we identified a corpus of 82 articles that explicitly employ complex network approaches to investigate a wide range of phenomena in science education research. Although these studies address diverse research questions and educational contexts, their use of network analysis reveals a limited number of recurring analytical logics. Specifically, the literature can be organized around four broad ways in which networks are mobilized: networks representing scientific concepts or ideas; social networks capturing patterns of interaction that shape educational processes; bibliographic networks derived from citation, co-authorship, or thematic relations among research outputs; and networks modeling attitudes, reasoning processes, or practices in learning and teaching. These categories are not intended as rigid or mutually exclusive—indeed, overlaps occur, such as bibliometric studies that also construct conceptual networks. Nevertheless, this classification emerged inductively from the reviewed corpus and reflects empirically grounded patterns in how network methods are currently used in science education research. As such, it provides a useful analytical framework for structuring the review and for highlighting both dominant approaches and underexplored connections across subfields.

\begin{figure}[ht]
    \centering
    \includegraphics[width=0.3\textwidth]{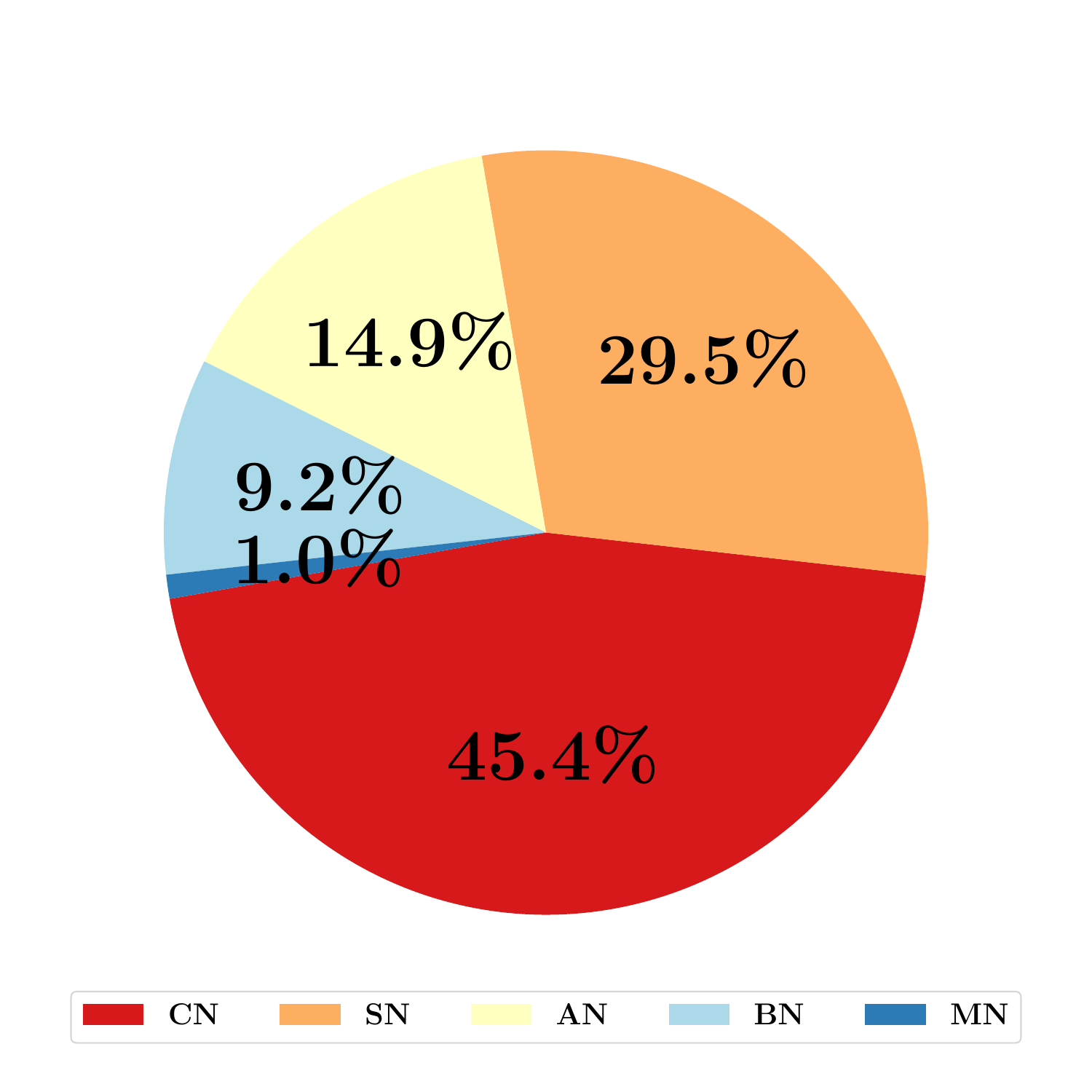}
    \caption{Frequency of articles in each category.}
    \label{fig:categories}
\end{figure}

The four main categories are named as concept networks (CN), which is further divided into three subcategories (concepts, surveys and other); social networks (SN), also with three subcategories (students, teachers and other); bibliographic networks (BN), with two subcategories (social and concepts); and attitudes networks (AN), with two subcategories (reasoning and practices). The distribution of the articles into these categories can be seen in Fig.~\ref{fig:categories} and in Tab.~\ref{tab:categories}.

\begin{figure}[ht]
    \centering
    \includegraphics[width=0.5\textwidth]{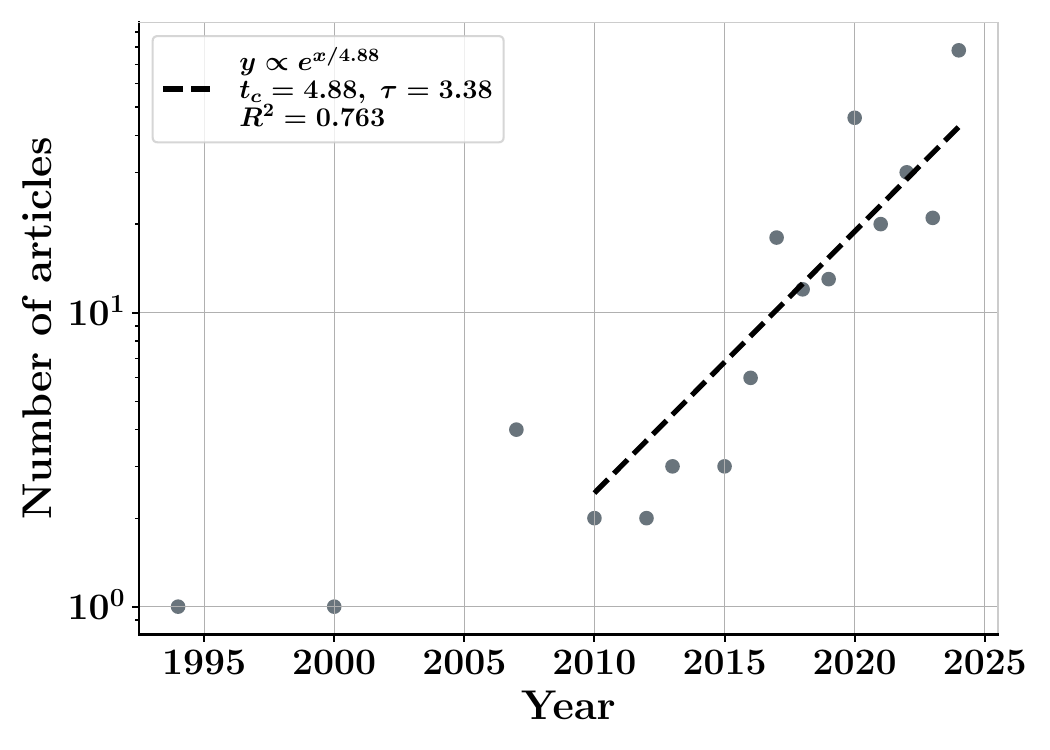}
    \caption{Temporal evolution of the number of articles. The line is an exponential fit to the total number of new articles per year with a doubling time of 3.38 years. }
    \label{fig:temporal}
\end{figure}

Regarding the temporal evolution of publications, Fig.~\ref{fig:temporal} shows a clear exponential increase in the number of articles employing network-based approaches in science education research, with an estimated doubling time of 3.38 years. Importantly, the growth curve displays no evidence of saturation, plateauing, or decline over the observed period. Such sustained exponential growth is characteristic of an emerging and consolidating research area, in which methodological adoption and conceptual exploration are still expanding rather than stabilising. This pattern suggests that network approaches are transitioning from niche or exploratory applications toward more mainstream methodological tools within the field. The absence of saturation further indicates that the diversity of educational contexts, data types, and theoretical frameworks amenable to network-based analysis has not yet been exhausted, pointing to substantial room for future development and innovation.

Before diving into the particularities of each category, let us here introduce a concise summary of them. 

\subsection{CONCEPT NETWORKS (CN)}
Studies on concept networks, in which the nodes represent concepts or ideas and the links typically encode epistemic relationships between them, although other ways of constructing networks are also found, such as based on frequency of occurrence or additional variables. This category is subdivided into three groups:
\begin{itemize}
\item \emph{ConceptNetwork-Concept (CN-C)} Studies that analyse networks of students’ concepts or ideas about a given topic, the evolution of these ideas over time, their clustering according to different variables, or comparisons of conceptual structures between students and teachers, among others.
\item \emph{ConceptNetwork-Surveys (CN-Su)} Studies that construct concept networks from responses to questionnaires or surveys, focusing on the analysis of survey items, response types, and the relationships between them.
\item \emph{ConceptNetwork-Other (CN-O)} Studies that employ concept networks related to the Nature of Science (NOS), typically using Epistemic Network Analysis (ENA). Note that other studies using ENA are not included here but rather in the first subcategory of this type, when they focus on concept networks about a specific scientific topic.
\end{itemize}

\subsection{SOCIAL NETWORKS (SN)}
In general, most articles in this category focus on student networks and examine their relationship with academic progress. To a lesser extent, teacher networks are analysed. This imbalance may be related to what is observed in another category—author or coauthorship networks that form research communities. However, in the present category, teacher networks are typically analysed in relation to attitudes toward students or issues more directly linked to classroom practice.
\begin{itemize}
\item \emph{SocialNetwork-Students (SN-S)} Student networks, addressing academic achievement, persistence, or trajectories within science pathways.
\item \emph{SocialNetwork-Teachers (SN-T)} Networks of teachers or researchers.
\item \emph{SocialNetwork-Other (SN-O)} Other examples of studies that connect individuals or social events, such as historical networks, where nodes may represent scientists or instruments/artifacts and links represent historical proximity or interaction.
\end{itemize}

\subsection{BIBLIOGRAPHIC NETWORKS (BN)}
\begin{itemize}
\item \emph{BibliographicNetwork-Social (BN-S)} Networks of people based on bibliographic data, such as coauthorship networks, used to study the temporal evolution of research communities. A particular case within this category includes a study that applies a similar approach to Twitter interactions, although the underlying idea is the same.
\item \emph{BibliographicNetwork-Concepts (BN-C)} Networks of concepts or topics constructed from keyword co-occurrence, co-citation, and/or coauthorship. Some studies in this subcategory also identify different ways of approaching the same research field (e.g., inquiry-based learning) and analyse trends or changes in the popularity of different topics over time.
\end{itemize}

\subsection{ATTITUDES NETWORKS (AN)}
Some studies use complex networks to analyse forms of argumentation, reasoning, or issues more closely related to cognitive psychology. Others focus on identifying social practices, attitudes, or learning difficulties, all within the context of science learning situations. Two subcategories can be distinguished:
\begin{itemize}
\item \emph{AbilitiesNetwork-Reasoning (AN-R)} Studies focusing on forms of argumentation, reasoning, understanding, and related cognitive processes.
\item \emph{AbilitiesNetwork-Practices (AN-P)} Studies related to practices, attitudes, behavioural patterns, and learning difficulties.
\end{itemize}

\begin{table}[ht]
\begin{tabular}{|c|c|c|c|}
\hline
Category & $n_p$ & Subcategory & $n_p$ \\
\hline

\multirow{3}{*}{Concept networks (CN)} & \multirow{3}{*}{39} & CN-C & 25 \\
                   &                     & CN-Su & 4 \\
                   &                     & CN-O & 10 \\
\hline

\multirow{3}{*}{Social networks (SN)} & \multirow{3}{*}{20} & SN-S & 17 \\
                   &                     & SN-T & 2 \\
                   &                     & SN-O & 1 \\
\hline

\multirow{2}{*}{Bibliographic networks (BN)} & \multirow{2}{*}{10} & BN-S & 3 \\
                   &                     & BN-C & 7 \\
\hline

\multirow{2}{*}{Abilities networks (AN)} & \multirow{2}{*}{13} & AN-R & 8 \\
                   &                     & AN-P & 5 \\
\hline
\end{tabular}
\caption{\label{tab:categories}Types of studies in science education that use complex networks according to the established coding scheme and the number of publications ($n_p$) identified.}
\end{table}

\section{Types of studies using complex networks in science education: review}

\subsection{Concept networks}

\subsubsection{Concept networks: concepts}

Within science education research, concept networks have been employed as powerful analytical tools to visualize and quantify how learners and teachers organize, relate, and transform scientific ideas. Studies in this category share a common goal: to move beyond traditional representations of conceptual understanding (e.g., concept maps or categorical coding) toward models that capture the relational structure and dynamics of knowledge. Nodes typically represent concepts, expressions, or epistemic elements, while links encode different forms of relationships—causal, epistemological, or co-occurrence patterns—thereby offering a systemic view of learning as a reorganization of conceptual connections rather than the mere acquisition of isolated facts. Along similar methodological lines, \cite{Siew2019} demonstrate how network-science measures applied to concept maps and semantic networks can capture individual differences in the organization of conceptual knowledge, showing that learners’ understanding can be meaningfully characterized by global properties such as network connectivity, clustering, and path structure rather than by isolated concept frequencies. In this vein, several recent contributions (e.g., \cite{Thurn2020,Kapuza2020,Podschuweit2020}) have leveraged network-theoretic metrics (such as centrality or community detection) applied to concept maps to capture structural growth, map density and concept centrality in domains including magnetism, electrostatics and statistical data analysis.

\begin{figure}[ht]
    \centering
    \includegraphics[width=0.4\textwidth]{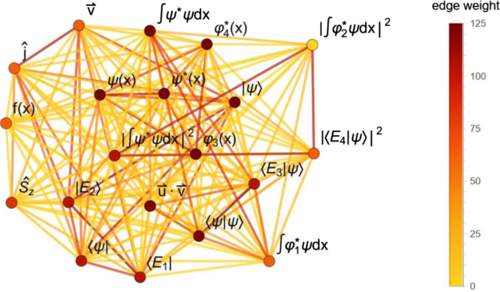}\\
    \includegraphics[width=0.3\textwidth]{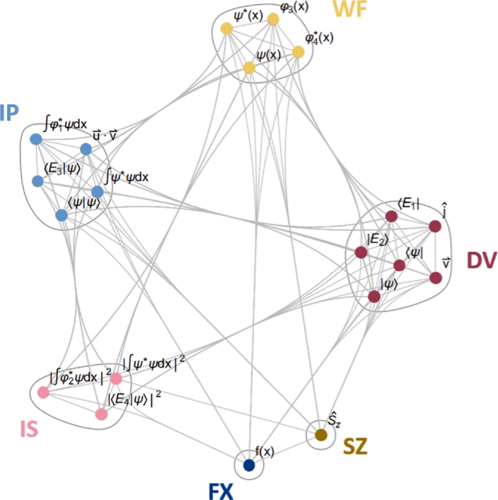}
    \caption{Examples of the networks extracted from \cite{Riihiluoma2024}: the upper panel shows an expression–concept network built from the responses of 139 students, where nodes represent survey expressions and edges indicate that two expressions were selected simultaneously for at least one concept (edge weights reflect the number of students). The lower panel shows the same network after community analysis, revealing six stable forms of grouping among the expressions..}
    \label{fig:example_concept_net}
\end{figure}

A first group of studies focuses on students’ evolving knowledge structures across disciplinary domains. For example, in the context of university quantum mechanics, \cite{Riihiluoma2024} model how students relate different mathematical expressions—such as Dirac notation and wave-function representations—by constructing networks where edges denote conceptual equivalence between symbolic forms. Through community-detection algorithms based on edge betweenness, the authors identify clusters reflecting how students implicitly group mathematically equivalent but epistemologically distinct expressions, revealing both coherent sub-systems of understanding and points of disconnection. Figure \ref{fig:example_concept_net} shows two examples of the types of networks used in this kind of study.
Similarly, in biology education, \cite{Todd2019,Todd2022} examine the transition from phenotypic to genotypic reasoning about genetics. By representing key ideas as nodes and causal relations as links, these studies show how students’ conceptual networks reorganize over time, initially fragmenting around surface-level attributes and later stabilizing into more mechanistic, molecularly grounded structures years after instruction—illustrating the long-term consolidation of conceptual coherence. 

Additionally, \cite{Nousiainen2020} analyzes pre-service teachers’ lexical networks related to wave–particle dualism. The results indicate that more diverse and coherently structured lexical networks are associated with a more multifaceted understanding of quantum entities, supporting the view of learning as the organization and refinement of knowledge networks.

Network approaches have also been used to compare taught and learned curricula. \cite{Lim2024} construct semantic networks of Mendelian inheritance concepts to contrast teachers’ intended structures with those reflected in students’ responses. Although the two sets of networks share similar nodes, the connectivity patterns differ: students’ networks exhibit weaker integration and fewer bridging concepts, emphasizing the importance of “conceptual bridges” to align instructional design with student cognition. Complementarily, \cite{Park2020} use network analysis of textual responses to conceptual physics questions to show how normative, mixed, and non-normative models of reasoning correspond to distinct network topologies—more integrated networks being associated with expert-like reasoning centered around core ideas. 

Another line of research explores the structure and coherence of disciplinary knowledge as manifested in teachers’ or experts’ representations. Classic works such as \cite{Koponen2010,Koponen2013} and \cite{Mäntylä2007} view scientific knowledge as a system of interconnected constructs where conceptual and methodological links jointly shape understanding. By comparing networks built by experts and preservice teachers, \cite{Koponen2010} demonstrate that experts’ networks are not only denser but also encode procedural and methodological relations (e.g., modeling, experimentation), suggesting that expertise involves both conceptual and epistemic integration. The pioneering study by \cite{Mäntylä2007} formalizes this insight by introducing graphical reconstructions (DRoP and DRoS) to represent how physical quantities emerge from measurement processes; nodes represent quantities, and edges depict experimental or theoretical dependencies. This early work framed conceptual development as the progressive interlinking of measurement, theory, and representation—a view later elaborated in \cite{Koponen2013}, which portrays conceptual change as the shift from “knowledge-as-pieces” to theory-like, coherent systems. More recently, \cite{Podschuweit2020} extend this approach by analyzing how pre-service or novice learners’ concept networks manifest transitional structures—less hierarchical, less dense and with fewer methodological links—highlighting how disciplinary coherence develops gradually and under guided instruction.

\cite{Turkkila2020} shows how social concept network analysis can be connected with social network analysis. Specifically, the study models students’ background knowledge as concept networks using centrality and similarity measures, and examines whether these knowledge structures are related to students’ participation and roles in online discussions, finding no direct relationship between the two.

Research on preservice teachers’ conceptual coherence further extends this tradition. \cite{Nousiainen2013} analyze concept maps of physics teachers-in-training, quantifying the epistemic quality of links according to criteria such as correctness and explanatory power. Even among advanced students, highly coherent networks are rare, underscoring the challenge of forming integrated knowledge systems. Similar concerns are revisited in \cite{Zhang2025} and \cite{Parrish2024}, who employ Epistemic Network Analysis (ENA) to map how teachers connect ideas while engaging in professional or reflective tasks. \cite{Zhang2025} examine teachers’ discourse during collaborative design of socioscientific issue lessons, showing how ENA captures shifts toward more balanced connections between content, pedagogy, and epistemic reasoning. \cite{Parrish2024}, in turn, combine ENA with a card-sorting methodology to model preservice teachers’ evolving conceptions of engineering, revealing a move from fragmented to expert-like networks that integrate problem definition, iteration, and evaluation criteria. 

\cite{Kubsch2020} shows how concept networks of students are correlated with their ability of student for transferring knowledge. Students with a more coherently organized knowledge network were more successful at knowledge transfer tasks.

ENA also underpins several recent studies exploring students’ conceptual integration during science learning. \cite{Li2024} combine natural language processing and ENA to model how middle-school students connect ideas about energy transfer in photosynthesis and respiration. Networks generated from students’ explanations illustrate that higher knowledge-integration scores correspond to denser, more mechanistically organized networks. Likewise, \cite{Peel2024} use ENA to trace pre-post changes in students’ reasoning about natural selection, showing that instruction promotes stronger integration among scientifically correct ideas and a decline in erroneous connections. The broader potential of ENA for science education is synthesized in the scoping review by \cite{Reid2024}, which surveys nineteen studies using this method. They conclude that ENA provides a versatile bridge between qualitative coding and quantitative modeling, particularly suited to visualizing complex patterns of conceptual and epistemic connections in discourse, writing, and assessment data.

Beyond individual cognition, concept network approaches have also been extended to curricular and affective domains. For instance, \cite{Caramaschi2022} apply ENA within the Family Resemblance Approach to map the representation of Nature of Science dimensions in the Italian physics curriculum. Their network analysis exposes missing epistemic and sociocultural connections—such as ethics and power structures—highlighting curricular imbalances and opportunities for reform. In a different context, \cite{Alo2020} use semantic networks to assess children’s conceptual associations with science and technology after informal workshops involving low-cost digital tools. Although the intervention elicited enthusiasm, the resulting networks showed persistent affective ambivalence and limited conceptual change, demonstrating how network mapping can reveal subtle attitudinal patterns. Finally, more theoretical contributions, such as \cite{Wu2022}, propose integrating complex network models with learning analytics and artificial intelligence to map and foster adaptive epistemic beliefs, pointing toward the future convergence of conceptual, computational, and educational network research. In line with this broader trend, \cite{Thurn2020} use longitudinal network analysis of students’ concept maps in electromagnetism to show that network centrality and connectivity increase over time, indicating progressive structuring of physics knowledge with instruction.

Historically, the use of network analysis to examine conceptual structures in science education can be traced back to early semantic network models such as \cite{Wilson1994}, which employed the Pathfinder algorithm to derive knowledge networks from students’ concept maps on chemical equilibrium. These early studies established methodological foundations later refined through more sophisticated approaches integrating graph-theoretical measures, ENA, and computational linguistics.

Taken together, the [CN-C] literature demonstrates a clear evolution—from static representations of concept maps to dynamic, data-rich models that capture both the structure and the process of conceptual development. Across topics ranging from quantum mechanics to biology and from teacher cognition to curriculum analysis, these studies converge on the view that understanding science learning requires tracing how ideas interconnect within epistemic systems. Concept networks thus serve not only as descriptive tools but as theoretical frameworks for reconceptualizing what it means to know, learn, and teach science as a complex, interconnected endeavor.

\subsubsection{Concept networks: surveys}

Within the category of Concept Networks Survey, researchers employ network analysis to model the structure of students’ responses to established diagnostic instruments and questionnaires. Rather than focusing on the explicit content of students’ conceptual explanations, these studies infer cognitive and epistemic structures from patterns of response co-occurrence, where nodes represent items or specific answer choices and edges quantify the degree of statistical association among them. This approach enables the detection of clusters of related ideas—whether correct, incorrect, or attitudinal—offering a relational perspective on assessment data that traditional psychometric tools cannot fully capture.

\begin{figure}[ht]
    \centering
    \includegraphics[width=0.5\textwidth]{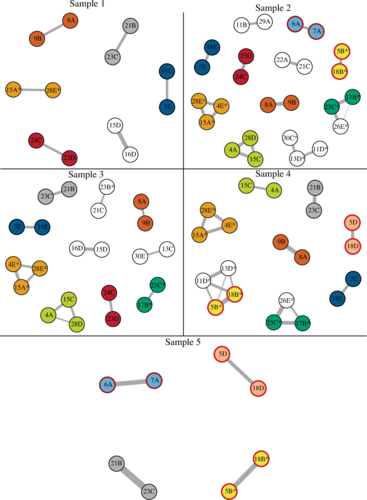}
    \caption{Network extracted from \cite{Wheatley2022}: Pretest partial correlation networks constructed from student responses to a conceptual survey. Communities indicate groups of answer patterns that tend to co-occur across students; edge width represents the strength of the association between responses. Correct answers are marked with an asterisk.}
    \label{example_survey_net}
\end{figure}

A central methodological innovation is the use of module analysis, a technique derived from community detection in complex networks, to identify groups of correlated responses. The foundational work of \cite{Wells2020,Wheatley2021,Wheatley2022} exemplifies this line of research. These studies analyze large datasets from canonical physics concept inventories—such as the Force and Motion Conceptual Evaluation (FMCE), the Force Concept Inventory (FCI), and the Conceptual Survey of Electricity and Magnetism (CSEM)—to reveal how students’ misconceptions cluster and evolve across instructional contexts. 
In \cite{Wells2020}, incorrect answers to FMCE items are treated as network nodes, linked according to the correlation in students’ selections across a sample of nearly 3,000 participants. The resulting networks expose communities of misconceptions corresponding to previously reported conceptual clusters (e.g., conflations between motion and force), while also uncovering gender-specific patterns and pre- to post-instructional changes. Building on this, \cite{Wheatley2022} extend module analysis to the FCI across five universities, finding that despite institutional variation, communities of misconceptions remain remarkably stable, indicating deeply entrenched alternative frameworks in students’ reasoning about Newtonian mechanics. Figure \ref{example_survey_net} shows an example of these networks extracted from \cite{Wheatley2022}.
\cite{Wheatley2021} apply a modified version of the method to the CSEM, introducing the notion of blocked-item communities—sets of interrelated items sharing contextual or conceptual dependencies. They demonstrate that certain items are conditionally dependent (e.g., a correct answer on one predicts correctness on the next), arguing that such dependencies should inform both the validation and rubric design of diagnostic assessments. Collectively, these studies show that network-based approaches can provide finer-grained insights into the latent conceptual architecture of misconceptions, bridging the gap between item-level statistics and cognitive modeling.

Parallel methodological developments appear in \cite{Dalka2022}, which proposes a network-analytic framework for Likert-style survey data. Here, items are treated as nodes and edges represent correlations in response patterns across participants. Community detection reveals clusters of items measuring related constructs, allowing researchers to diagnose redundancy and assess construct validity without relying solely on factor analysis or principal component analysis (PCA). The authors argue that modularity analysis provides a deeper and more interpretable view of the structure of survey data, capable of capturing nonlinear relationships among items that PCA may obscure. This study thus broadens the application of network methods from conceptual inventories to attitudinal and affective instruments within physics education research.

Taken together, these works redefine the use of diagnostic questionnaires in science education. Traditional psychometrics typically focus on individual item difficulty or latent traits, whereas network analysis treats the entire set of responses as an interconnected system, enabling the identification of emergent conceptual communities and cross-item dependencies. By visualizing and quantifying these relations, researchers can trace the collective topology of student reasoning, identify clusters of mutually reinforcing misconceptions, and refine the design and interpretation of assessment tools.

In summary, [CN-Su] studies demonstrate that applying complex network methods to survey data provides a powerful complement to classical statistical approaches. Whether through module analysis of physics inventories \cite{Wells2020,Wheatley2021,Wheatley2022}, correlation networks of Likert-style items \cite{Dalka2022},  
these contributions collectively advance a more structural understanding of students’ knowledge organization. Beyond validation, they reveal how conceptual coherence and error persistence manifest not only in students’ discourse but in the statistical patterns of their responses, positioning network analysis as a key methodological bridge between cognitive theory, assessment design, and learning analytics in science education.

\subsubsection{Concept networks: others}

Research employing concept networks to explore the Nature of Science (NOS) has expanded notably in recent years, particularly through the adoption of Epistemic Network Analysis (ENA) as a visual and quantitative tool to map how NOS dimensions co-occur in curricula, assessments, and participants’ conceptions. Unlike studies that model conceptual understanding of scientific content (e.g., energy, genetics), this line of work examines how learners, teachers, and institutional documents represent what science is—its epistemic, cognitive, and social dimensions. Although the methodological sophistication of ENA is often foregrounded, many of these studies converge on a relatively simple analytical goal: to reveal whether and how NOS elements are interconnected, and to what extent educational systems emphasize the social-institutional side of science alongside its cognitive-epistemic core.

The most visible trend in this corpus is the use of ENA combined with the Family Resemblance Approach (FRA) or its reconceptualized version (RFN) as the theoretical backbone for curriculum and textbook analyses. Studies such as \cite{Xie2025,Gao2025,Pimenta2025,Han2024,Cheung2020} apply ENA to national standards and curricular frameworks from China, Brazil, and Hong Kong. Across these contexts, nodes represent NOS categories—typically divided between cognitive-epistemic aspects (e.g., scientific practices, theory formation, methods) and social-institutional ones (e.g., professional organizations, ethics, cultural and financial systems)—while edges correspond to their co-occurrence within curricular statements or textbook sections.
Consistently, these analyses find asymmetric network structures: cognitive-epistemic elements are densely connected, forming the dominant core of the curriculum, whereas social-institutional aspects appear peripheral and weakly integrated. For example, in the Brazilian BNCC, \cite{Pimenta2025} report strong interlinking between “scientific practices” and “scientific knowledge” but sparse connections to social values and organizational dimensions. Likewise, the longitudinal study of Chinese physics curricula by \cite{Xie2025} reveals only marginal growth in social NOS connectivity from 2001 to 2022, and \cite{Gao2025} find similar patterns in biology, indicating the persistence of an epistemically narrow portrayal of science. Even in textbooks—where integration might be more flexible—\cite{Han2024} note that quantum physics materials present a relatively balanced yet still cognitively centered picture, with the Nature of Scientific Theories acting as the main connective hub, while Social and Cultural Embeddedness remains underrepresented.

Beyond curriculum analysis, ENA has been used to investigate learners’ and teachers’ conceptions of NOS through classic assessments such as the Views of Nature of Science (VNOS-B). Early work by \cite{Peters-Burton2015} applied ENA to middle-school students’ pre- and post-responses, showing that instruction leads to denser, more integrated NOS networks—evidence of conceptual growth that traditional VNOS scoring schemes might overlook. Later, \cite{Mulvey2021} introduced individual Epistemic Network Analysis (iENA) to examine graduate science teachers’ responses, capturing nuanced shifts in how participants connected empiricism, theory, and creativity. These findings demonstrate that ENA can quantify structural changes in NOS understanding at the individual level, bridging qualitative and quantitative approaches. Extending this methodology, \cite{Peters-Burton2019} reanalyzed VNOS data to highlight how ENA surfaces clusters of related NOS ideas and reveals missing conceptual connections (e.g., naïve empiricism before instruction).

Recent comparative and re-analytical efforts push the approach toward a cross-population perspective. In \cite{Peters-Burton2023}, ENA is used to contrast the conceptual networks of scientists, teachers, and students. Predictably, scientists’ networks are denser and more coherent, whereas those of students and teachers remain fragmented, underscoring the epistemic gap between professional and educational representations of science.

At the meta-analytic level, \cite{Cheung2023} conduct a systematic review of studies grounded in the Family Resemblance Approach and employ ENA to visualize how FRA categories have been connected across 32 empirical papers. The review itself reproduces the same structural imbalance observed in curriculum studies: tightly knit cognitive-epistemic clusters contrasted with sparse, peripheral social-institutional links. The authors argue that this consistent asymmetry reflects not only curricular design but also the research community’s conceptual bias toward the epistemic core of science, urging future studies to strengthen integration of the sociocultural and institutional dimensions that shape scientific practice.

Taken together, these NOS-related studies show a coherent, if somewhat repetitive, methodological pattern: ENA serves primarily as a confirmatory visualization tool rather than a hypothesis-generating analytic engine. While the method successfully illustrates patterns of co-occurrence and allows for appealing graphical representation, its interpretive depth is often limited to reaffirming what qualitative content analysis already indicates—namely, that cognitive-epistemic dimensions dominate most representations of science in curricula, textbooks, and learners’ thinking. Nevertheless, ENA contributes a valuable structural layer to NOS research by making visible the degree of integration (or disconnection) among epistemic and social components of scientific understanding.

Overall, the [CN-O] body of work highlights both the promise and current limitations of using complex network approaches in meta-scientific contexts. On the one hand, ENA provides a rigorous, replicable framework for examining how NOS ideas are distributed and interconnected across educational artifacts and populations. On the other hand, the findings reveal a field still seeking theoretical depth and diversity of application. To advance beyond descriptive mapping, future research might combine ENA with discourse-based or cognitive-modelling approaches, enabling analyses that capture not just what NOS elements co-occur, but how and why they are linked in learners’ epistemic reasoning about science.

\subsection{Social networks}

\subsubsection{Social networks: students}

The application of social network analysis (SNA) to science education has yielded a rich body of research focused on how students’ patterns of social connection shape their engagement, achievement, and persistence in science. Rather than treating learning as an individual process, these studies conceptualize it as a socially distributed phenomenon, emphasizing how positions within classroom or extracurricular networks influence both academic outcomes and affective dimensions such as science identity or commitment to STEM pathways. Across contexts—from middle school clubs to university physics courses—network measures such as centrality, density, modularity, and tie strength serve as proxies for students’ levels of participation and integration in the learning community.

\begin{figure}[ht]
    \centering
    \includegraphics[width=0.5\textwidth]{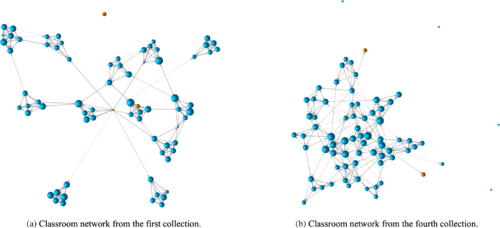}
    \caption{Network extracted from \cite{Williams2019}: Evolution of a student interaction network at two time points during a course. Early ties largely reflect seating proximity, while later the network shows a more complex interaction structure, illustrating its development over time. Node size represents closeness centrality at the later measurement; highlighted nodes correspond to students who did not complete the course.}
    \label{example_social_student_net}
\end{figure}

A core line of work, originating in Physics Education Research (PER), systematically relates students’ network positions to academic performance and persistence. Foundational studies by \cite{Zwolak2017,Zwolak2018,Vargas2018,Williams2019,Traxler2018} constructed classroom networks using survey data where students reported peers with whom they interacted, collaborated, or sought help. Network metrics—particularly degree and betweenness centrality—were then correlated with course grades, persistence, and retention. These studies converge on a consistent pattern: students more centrally embedded within classroom networks tend to achieve higher grades and show greater persistence in STEM pathways. For instance, \cite{Vargas2018} find that students who collaborate frequently and with a diverse set of peers outperform those with narrower collaborations, suggesting that network diversity enhances access to cognitive resources. Similarly, \cite{Williams2019} demonstrate that engagement and academic success co-evolve: as classroom interactions become more cohesive over time, average achievement increases, indicating a reciprocal reinforcement between community structure and learning. This is shown in Figure \ref{example_social_student_net}, extracted from \cite{Williams2019}.
Complementary work by \cite{Zwolak2017,Zwolak2018} extends these insights to persistence in introductory physics, distinguishing between in-class and out-of-class networks. Their findings reveal that while high and low achievers’ persistence depends primarily on grades, for mid-performing students, social connectivity acts as a decisive factor—a result that underscores the role of community integration as a mediating mechanism in STEM retention.

In addition to retention and performance, network position has also been linked to affective outcomes: \cite{Dou2016} use SNA to examine how students’ centrality metrics (e.g., inDegree, outDegree, PageRank) predict changes in physics self-efficacy over the course of an interactive introductory physics course, finding that higher centrality correlates with stronger maintenance (or smaller decrease) of self-efficacy.
This analysis is extended \cite{Dou2019} to multiple affective constructs, showing that students’ positions in classroom interaction networks are associated not only with self-efficacy but also with physics identity and interest, and that these relationships evolve over time as students’ social embeddedness in the course develops.

More recent studies expand the scope of SNA to address new learning contexts, particularly online and hybrid environments. During and after the COVID-19 pandemic, researchers examined how the shift in modalities affected network formation and its link to achievement. \cite{Crossette2023} compare collaboration networks across in-person, online, and hybrid physics courses, finding that although modality alone does not predict performance, missing or incomplete network data (absent nodes) significantly weakens observed correlations, revealing the methodological sensitivity of network measures to participation loss. Likewise, \cite{Pulgar2022,Pulgar2023} use network data from Chilean high school physics classes to explore how the nature of relationships—friendship ties, perceived prestige, or collaborative interactions—modulates learning outcomes across remote and hybrid settings. Their results show that friendship-based collaborations enhance academic gains, whereas prestige-based ties have neutral or even diminishing returns. Moreover, in hybrid environments students display richer and more diverse connections, while in fully online settings academic performance becomes a dominant organizing principle, leading to more hierarchical, less diverse networks.

A complementary recent contribution, \cite{Rosen2022}, explores student interaction networks in a laboratory-based physics setting, demonstrating that network centrality and participation patterns continue to predict course engagement and lab performance.

These findings align with broader efforts to understand the structure and evolution of learning communities in active-learning and digital settings. \cite{Traxler2020} examine network positions across a semester in an active-learning physics course, identifying a trend toward increased network density and modular stability: students’ positions become more consistent and cohesive over time, with most participants gravitating toward stable subgroups. This temporal dimension reveals how community structures consolidate as courses progress, and how integration into the “giant component” correlates with engagement and persistence.

The literature also includes attempts to diversify SNA methodologies within science education. For example, \cite{Traxler2024} advocate for the adoption of ego-centered and qualitative network approaches that foreground the interpretive richness of individual connections. By situating network data in narrative or ethnographic contexts, they argue, researchers can capture nuances of identity formation, motivation, and belonging that quantitative metrics alone overlook. Similarly, \cite{Swain2022} extend SNA to online ecology courses, showing how peer interactions influence group formation, topic selection, and research performance, highlighting that social structure shapes not only outcomes but also the direction of inquiry in project-based learning environments.

Beyond academic outcomes, social network perspectives have been employed to explore affective and identity-related constructs in science learning. \cite{Hill2024} use SNA to investigate how participation in extracurricular science clubs—and the friendship networks that bridge participants and non-participants—affect students’ science identity. They find that both direct and indirect ties to club members foster stronger identification with science, illustrating how the social reach of science engagement extends beyond formal participation. Likewise, \cite{Patchen2015} model classroom collaboration networks in high school chemistry, showing that collective structures of interaction contribute to broader perceptions of what “doing science” entails, linking collaboration to epistemological and motivational growth.

At higher educational levels, network approaches are being used to analyze research environments and professional learning communities. For instance, \cite{Jones2025} track the evolution of social connections among graduate students and faculty involved in convergent, interdisciplinary research centers. Over the course of an academic year, participants’ networks become more integrated and cross-institutional, indicating that engagement in convergent science environments fosters intergroup connections and epistemic diversity. 

Collectively, the [SN-S] literature portrays learning in science not merely as an individual cognitive achievement but as a socially situated process of networked participation. Across studies, three key insights emerge:

\begin{itemize}
\item Structural integration predicts success. Students who are more central, connected, or bridge otherwise separated groups tend to perform better and persist longer in STEM.

\item Context matters. Network structure varies across modalities (in-person, online, hybrid) and institutional settings, shaping opportunities for collaboration and equity.

\item Methodological diversification is underway. From quantitative centrality analyses to qualitative ego-network approaches and mixed methods linking social and epistemic dimensions.
\end{itemize}

Together, these studies demonstrate that network analysis provides not only diagnostic insight into classroom dynamics but also a conceptual framework for understanding how social structure underpins learning, identity, and persistence in science education.

\subsubsection{Social networks: teachers}

Studies categorized under [SN-T] extend social network analysis (SNA) to the professional and institutional contexts of science and STEM teachers, focusing on how educators’ social positioning, beliefs, and communication patterns shape their pedagogical practices and collaborative dynamics. While research on student networks tends to link social connectedness to achievement or persistence, this smaller body of work examines how teachers’ interactions and value systems—conceptualized as networked phenomena—mediate instructional innovation, interdisciplinary work, and the diffusion of pedagogical ideas within educational communities.

A classical contribution to this line of inquiry is the study by \cite{Judson2007}, which investigates the role of constructivist-oriented teachers within faculty communication networks in biology and mathematics departments. Using sociograms to map the frequency and type of collegial interactions, the authors show that constructivist teachers occupy more central and influential positions in pedagogical conversations than their more traditional counterparts. These teachers act as bridging agents who connect otherwise disconnected colleagues, serving as conduits for the exchange of teaching strategies, content-specific discussions, and reform-oriented ideas. The study thus frames the implementation of constructivist pedagogy not only as an individual cognitive stance but as a network-mediated process dependent on social capital and communicative embeddedness within the faculty.

More recently, network approaches have been combined with psychometric and sociocultural frameworks to explore the underlying values that influence teacher collaboration. \cite{Yang2024} exemplifies this trend by examining STEM teachers’ power distance values (PDVs)—beliefs about hierarchy, authority, and social distance in professional relationships—within interdisciplinary teams. Here, nodes represent specific PD dimensions (e.g., hierarchy acceptance, prestige orientation, relational distance), while edges capture statistical associations between them as derived from confirmatory factor analysis. The resulting network visualizations identify which dimensions are most central to teachers’ value systems and how clusters of beliefs interact to shape collaboration patterns. Findings indicate that hierarchical and authority-based orientations form the core structural components of PDVs that influence how teachers navigate interdisciplinary partnerships. By integrating CFA and SNA, this study provides a methodological bridge between latent-variable modeling and network theory, yielding a more nuanced, systemic view of how cultural values are organized and expressed in collaborative behavior.

Together, these works demonstrate the potential of social network perspectives to illuminate the relational infrastructure of teaching communities in science education. They reveal that both pedagogical innovation and interdisciplinary collaboration are not solely outcomes of individual expertise but are deeply embedded in the networked ecology of professional interaction. Although this subfield remains relatively small, its methodological diversity—ranging from sociograms of communication to hybrid statistical-network models—points toward future opportunities to link teachers’ epistemic beliefs, social positioning, and institutional context within integrated frameworks of networked professional learning.

\subsubsection{Social networks: others}

The [SN-O] category extends the scope of social network analysis in science education beyond contemporary classroom or professional interactions to include historical and sociocultural dimensions of scientific practice. In this subfield, network methods are used not only to model social relations among learners or teachers but also to help students reconstruct and interrogate the history of science itself through the lens of connectedness and exchange.

A representative example is provided by \cite{Alcantara2020}, who introduce historical network analysis as a pedagogical strategy for engaging students with the history and nature of science. In this study, high school students explore archival correspondence and data about historical figures in science, using digital humanities tools such as ePistolarium to visualize networks of communication among scientists and the material culture of their work. Here, nodes represent both human actors (e.g., scientists) and non-human entities (e.g., instruments, letters, or research artifacts), while edges denote documented interactions or exchanges. This networked reconstruction allows learners to perceive scientific activity as a web of relationships unfolding across space and time, emphasizing collaboration, communication, and the socio-material context of discovery.

Although still rare within the broader literature, this work illustrates how network analysis can bridge history of science education and data-informed inquiry. By positioning learners as analysts of real historical networks, rather than passive recipients of scientific milestones, such approaches cultivate meta-scientific reflection and foster a deeper appreciation of the complex social dynamics that underpin scientific knowledge production. In this way, the [SN-O] category contributes to expanding the conceptual reach of network thinking in science education—from modeling current social systems to reconstructing the historical architectures of science itself.

\subsection{Bibliographic networks}

\subsubsection{Bibliographic analysis: social networks}

Research in the [BN-S] category applies social network analysis to the scholarly and professional communities of science education, uncovering how patterns of collaboration, communication, and co-authorship shape the field’s intellectual and social evolution. Here, nodes typically represent individual researchers, authors, or professional participants (e.g., teachers and administrators on social media), and edges correspond to forms of scholarly interaction—coauthorship, citation, or dialogic exchange. These studies situate science education as a complex, evolving social system, where the structure of connections influences not only knowledge dissemination but also the formation of subcommunities, thematic specializations, and opportunities for innovation.

\begin{figure}[ht]
    \centering
    \includegraphics[width=0.3\textwidth]{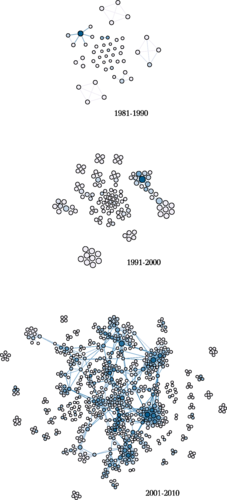}
    \caption{Network extracted from \cite{Anderson2017}: Evolution of a scientific collaboration network across three decades (1980s–2000s), illustrating changes in the structure of co-authorship relationships over time.}
    \label{example_biblio_social_net}
\end{figure}

A seminal example is provided by \cite{Anderson2017}, who examine the coauthorship network in physics education research (PER) over three decades across The American Journal of Physics, PERC Proceedings, and Physical Review Physics Education Research. By mapping collaboration patterns among hundreds of authors, the study captures the institutional maturation of PER from a loosely connected collection of individual scholars to a structured research community. Two key structural shifts are identified: first, a discrete rise in coauthorship coinciding with the establishment of dedicated PER conferences, signaling the institutional consolidation of the field; and second, a behavioral change among established authors, who move from collaborating with newcomers (“outsiders”) toward forming tighter intra-core collaborations. This evolution—from peripherally connected to densely clustered—illustrates how social and institutional mechanisms drive the self-organization of a research field, mirroring broader dynamics of disciplinary professionalization. An example of this kind of networks is shown in Figure \ref{example_biblio_social_net}.
Expanding this perspective beyond a single subdiscipline, \cite{Fontes2025} analyze the global coauthorship network of science education research grounded in Cultural-Historical Activity Theory (CHAT). Their findings reveal a highly fragmented topology, characterized by small, cohesive clusters with limited cross-cluster collaboration. Although some small-world properties emerge (short average path lengths, local clustering), the absence of strong interconnections between clusters constrains theoretical integration and interdisciplinary dialogue. The authors interpret this fragmentation as symptomatic of a “silo effect” in which researchers within similar theoretical traditions preferentially collaborate, thereby reinforcing conceptual isolation. They advocate for the creation of cross-thematic and international collaborations to facilitate knowledge integration and enhance the field’s collective development.

A complementary but distinct contribution comes from \cite{Rosenberg2020}, who move from formal publication-based networks to digital professional networks, analyzing the NGSSchat community on Twitter. This study examines over 7,000 tweets to map interactions among teachers, administrators, and researchers discussing the Next Generation Science Standards (NGSS). Using SNA, they identify key actors (brokers, hubs, peripheral participants) and explore how communication patterns sustain engagement and professional learning online. The findings reveal a multi-layered, participatory professional ecology, where educators and researchers co-construct knowledge and share resources, but also where influence and information flow are unevenly distributed. Importantly, the study highlights how social media-based networks can extend traditional professional learning communities, providing new channels for informal collaboration and reform-oriented discourse in science education.

Taken together, these studies demonstrate that the social structure of science education research and practice mirrors the same principles of complex network organization observed in scientific disciplines themselves. Coauthorship and digital communication networks both exhibit clustering, preferential attachment, and core-periphery dynamics, reflecting how collaboration norms and institutional structures evolve over time. The [BN-S] literature thus contributes to a meta-perspective on the field: by examining how science educators connect, publish, and communicate, it exposes the social architecture of knowledge production that underpins the very research on teaching and learning in science.

\subsubsection{Bibliographic analysis: concept networks}

Research in the [BN-C] category employs bibliometric and concept-based network analyses to chart the intellectual landscape of science education. In these studies, nodes typically represent keywords, authors, or cited references, and edges correspond to co-occurrence or co-citation relationships. By applying graph-theoretical and clustering algorithms, researchers uncover thematic communities, identify emerging “hot topics,” and trace the evolution of conceptual linkages across decades of scholarship. These works collectively transform the vast corpus of science education literature into a structured knowledge network, revealing how ideas, methods, and topics coalesce and diverge over time.

\begin{figure}[ht]
    \centering
    \includegraphics[trim={6cm 0cm 0cm 0cm}, clip,width=0.45\textwidth]{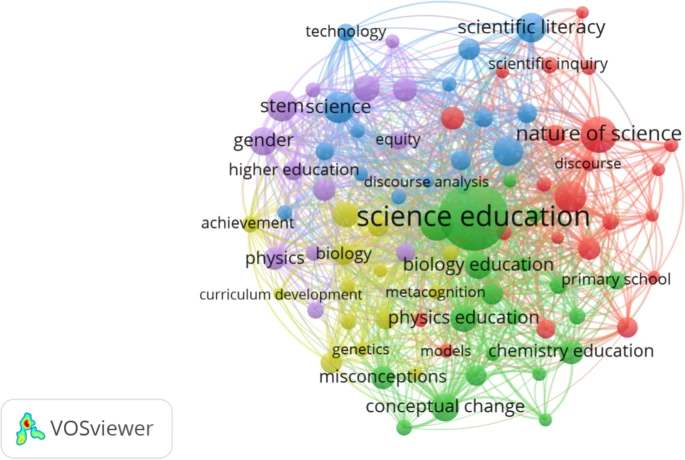}
    \caption{Network extracted from \cite{Tosun2024}: Keyword co-occurrence network constructed from Science Education Research (SER) articles, highlighting the most frequently used terms and their relationships across the literature.}
    \label{example_biblio_concept_net}
\end{figure}

One of the most comprehensive contributions in this line is the large-scale bibliometric mapping by \cite{Wang2023}. Analyzing over 6,200 articles published between 2001 and 2020 in seven SSCI-indexed journals, the authors construct multiple networks—co-authorship, co-citation, and co-occurrence—to identify key research themes and their chronological transitions. Their results delineate three developmental phases: an early focus on learning and identity in informal contexts; a mid-period dominated by scientific literacy, social issues, and formal schooling; and a recent shift toward argumentation, STEM integration, and methodological reflection. Collaboration networks reveal strong international clustering across North America, Europe, and Oceania, and the most influential nodes correspond to methodological and reform-oriented topics such as standards, inquiry, and teacher professional development. Similarly, \cite{Tosun2024} extend the scope to 13,000+ articles spanning four decades, using VOSviewer to visualize co-occurrence and co-citation networks. Their findings confirm a steady expansion of the field since 2007 and highlight recurring conceptual anchors—“science education,” “argumentation,” “scientific literacy,” “conceptual change,” and “nature of science”—as persistent hubs structuring research discourse.
An example of a network extracted from \cite{Tosun2024} is shown in Figure \ref{example_biblio_concept_net}.

Complementary studies zoom into specific thematic niches or disciplinary lenses. \cite{Giamellaro25} conduct a bibliometric network review of contextualized science learning, analyzing how content-context relationships are conceptualized across the literature. Thirteen distinct research communities emerge, each representing a different approach to contextualization—ranging from situative and socio-cultural perspectives to cognitive and curricular ones. This mapping clarifies that “contextualization” operates as a multi-dimensional construct rather than a unified concept, helping delineate boundaries and overlaps among theoretical traditions. Similarly, \cite{Tang2024} examine the evolution of socioscientific issues (SSI) research through co-word network analysis of 334 papers (2000–2021). Their networks highlight thematic clusters around argumentation, decision-making, and scientific literacy, and reveal growing diversification in how SSI is framed, suggesting that this subfield is shifting toward more integrative, justice-oriented perspectives.

Thematic co-word analyses have also been used to track macro-trends in STEM education. \cite{Hsu2024} apply co-word and cluster analysis to 1,248 STEM education articles (2011–2020), identifying three major thematic constellations—STEM integration, equity, and digital learning. Temporal mapping of co-occurrence strength reveals dynamic topic trajectories, showing, for instance, a surge of interest in computational and design-based learning alongside declining attention to early integration frameworks. Such studies exemplify the potential of bibliometric networks to detect emergent paradigms and thematic transitions in real time. Similarly, \cite{Jia2024} narrow the focus to the rise of artificial intelligence (AI) in science education (2013–2023). Their CiteSpace-based cluster analyses visualize a small but rapidly expanding community, linking AI research to subthemes such as adaptive assessment, learning analytics, and personalized feedback—topics that are beginning to reshape the methodological toolkit of science education research itself.

In addition to co-word and co-citation mapping, some studies incorporate Epistemic Network Analysis (ENA) to conceptualize literature reviews as relational models of research constructs. \cite{Vo2024}, for example, analyze 63 empirical studies on scientific inquiry published between 2020 and 2024. Using ENA, they model the co-occurrence of inquiry components—hypothesis generation, experimental design, data analysis, and conclusion drawing—across the reviewed papers. The resulting networks reveal that data analysis serves as a central hub connecting procedural and conceptual aspects of inquiry, suggesting that empirical research increasingly emphasizes data-driven reasoning as the nexus of scientific practice. This application of ENA demonstrates how network-analytic methods can enrich systematic reviews by visualizing conceptual structures within the research corpus itself.

Across all [BN-C] studies, common patterns emerge:

\begin{itemize}
\item Convergence around key epistemic anchors (scientific literacy, argumentation, inquiry, STEM integration) that define the enduring thematic spine of science education.

\item Expansion of network-analytic techniques, from traditional co-word and co-citation mapping to hybrid models integrating ENA and machine-learning-assisted bibliometrics.

\item Shift from static description to dynamic modeling, where time-sliced and longitudinal analyses reveal how topics rise, merge, or fragment in response to educational reforms and global trends.
\end{itemize}

Together, these works portray the science education research community as a complex, evolving knowledge ecosystem, whose conceptual architecture can be quantitatively mapped and visually explored. The [BN-C] category thus exemplifies how complex-network methodologies can move beyond descriptive bibliometrics toward cognitive cartography—a network-based understanding of how ideas in science education connect, evolve, and reorganize over time.

\subsection{Abilities networks}

\subsubsection{Abilities networks: reasoning}

The [AN-R] category encompasses studies that employ network approaches to model scientific reasoning, argumentation, and cognitive organization during learning processes. Across these works, networks are used not merely as analytical tools but as representations of thinking itself—whether conceptual, neural, or epistemic. The diversity of approaches within this subcategory underscores the flexibility of network science as a bridge between cognitive, social, and computational views of reasoning in science education.

\begin{figure}[ht]
    \centering
    \includegraphics[width=0.45\textwidth]{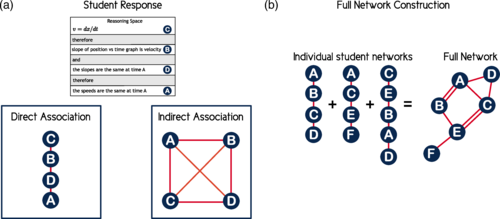}
    \caption{Network extracted from \cite{Speirs2024}: Reasoning network constructed from students’ inferential chains, where nodes represent elements of reasoning and edges indicate how these elements are connected when students build explanations. The network structure reveals common pathways in students’ reasoning.}
    \label{example_abi_reasoning_net}
\end{figure}

At the conceptual level, several studies visualize the structure of students’ reasoning chains or scientific explanations as explicit concept networks. For example, \cite{Speirs2024} propose a network-analytic framework to explore students’ qualitative inferential reasoning chains. Students are given sets of scientifically valid statements about a phenomenon and asked to link them logically. The resulting networks—where nodes represent statements and edges the inferential links students construct—reveal differences in the coherence, complexity, and integration of reasoning paths. The method not only captures how learners organize causal and explanatory relationships but also provides feedback for improving task design and scaffolding reasoning skills. Figure \ref{example_abi_reasoning_net} shows an example of this kind of networks.
Likewise, \cite{Wagner2023} employ concept-mapping techniques combined with network metrics (e.g., size, density, connectivity) to assess the quality of scientific explanations in student writing. Comparing student and expert networks, they show that expert explanations exhibit denser and more hierarchically organized structures, emphasizing that the form of reasoning can be quantitatively characterized through network topology. 

A further extension of network-based analyses is offered by \cite{Moreno-Esteva2020}, who apply a directed graph model to gaze-tracking data in the context of biodiversity education. By representing visual transitions between areas of interest as interconnected nodes, they characterize individual scanning signatures and show how differences in network structure reflect varying observational competencies between experts and novices.

At a more socio-interactional scale, \cite{Gonzalez19} use social network analysis (SNA) to examine the interactional dimension of argumentation in middle-school science classrooms. Here, the networks do not represent ideas but participants—the social fabric through which reasoning unfolds. By visualizing who speaks to whom and how often during argumentation sessions, the authors reveal distinct participation patterns that shape the flow of reasoning. Their work highlights that productive argumentation is as much a socially distributed process as a cognitive one, and that SNA can uncover the dynamics of epistemic agency in dialogic learning environments.

Other contributions extend the notion of reasoning networks to the semantic and affective domains of scientific understanding. \cite{Pietrocola2025} use semantic network analysis to explore how students conceptualize risk and uncertainty when reasoning about human-related hazards. In these networks, concepts such as “risk,” “probability,” and “responsibility” function as hubs whose interconnections reveal the epistemic and moral frameworks students invoke under uncertainty. By quantifying the semantic cohesion of these networks, the study demonstrates how reasoning in ill-defined, socio-scientific contexts depends on linking cognitive and affective dimensions.

A distinctive branch of this subfield focuses on neural and epistemic representations of reasoning. \cite{Lee2012} report one of the earliest attempts to link cognitive training in hypothesis generation and understanding with neural connectivity patterns observed through fMRI. Their results show distinct yet trainable functional network systems underpinning these two forms of scientific reasoning—generative versus interpretive—thus providing biological evidence that reasoning abilities correspond to separable but interrelated neural architectures. In a similar systems-level spirit, \cite{Sung2024} employ Epistemic Network Analysis (ENA) and multimodal learning traces to study self-regulated learning (SRL) during biology laboratory tasks. By modeling how SRL behaviors (planning, monitoring, reflection) co-occur over time, they reveal distinct temporal network patterns between high- and low-performing students. ENA visualizations show that effective learners exhibit denser, temporally stable connections among SRL components, indicating that expert-like regulation involves tighter coupling of cognitive and metacognitive actions.

Finally, theoretical work such as \cite{Roth2000} uses artificial neural networks (ANNs) as analogical and computational models of learning and knowing in science. Building on Piagetian problems of cognitive equilibrium, Roth argues that ANN models—due to their interactive and adaptive nature—capture the emergent, self-organizing features of human reasoning better than linear or symbolic accounts. Though predating modern learning analytics, this study anticipates later developments that conceptualize reasoning as a dynamic networked process both cognitively and computationally.

Taken together, the [AN-R] corpus illustrates a conceptual shift from viewing reasoning as a linear sequence of propositions to understanding it as a complex, interconnected system. Whether modeled through student-constructed concept maps, neural connectivity, or epistemic co-occurrence, these studies converge on the insight that reasoning proficiency depends on the structure and integration of connections—between ideas, between learners, and between cognitive and metacognitive functions. By foregrounding network topology as a marker of cognitive sophistication, this line of research opens powerful methodological pathways for quantifying and visualizing the architecture of scientific thinking.

\subsubsection{Abilities networks: practices}

The [AN-P] subcategory comprises studies that use complex network methods—particularly ENA—to investigate how practices and interactions are structured within teaching and learning processes. Rather than focusing on conceptual reasoning per se, these works model the behavioral and communicative fabric of educational episodes, highlighting how teachers, tutors, and students coordinate social, linguistic, and pedagogical actions in real time. Through this lens, networks become a way to visualize and quantify practice-based cognition, revealing how professional and interpersonal competencies manifest through patterns of co-occurring behaviors.

One of the most detailed applications of this approach is presented by \cite{Olsen2023}, who study social behavior patterns in teaching assistant–student interactions during online physics courses. Combining qualitative coding with a modified form of ENA, the authors model instructors’ social practices—such as positive tone, empathy, or neutral feedback—as nodes, and their co-occurrences within conversational “stanzas” as weighted links. The resulting networks show that high-rated instructors engage in denser and more balanced configurations of positive and empathetic behaviors, whereas low-rated ones rely more heavily on negative or corrective patterns. This nuanced representation of pedagogical discourse as a behavioral network offers insights into how affective and communicative dimensions shape students’ perceptions of teaching quality, and illustrates ENA’s adaptability beyond purely cognitive analyses.

At a broader scale of instructional interaction, \cite{Wang2024} apply ENA and complementary computational techniques (AI-based dialogue analysis and latent semantic analysis) to 1,500 online one-on-one tutoring sessions across different educational levels (primary to high school). Here, networks capture co-occurrences among tutoring strategies, question types, and feedback patterns. Distinct configurations emerge for each educational level: tutors working with younger students exhibit more tightly interconnected patterns of scaffolding and encouragement, whereas those with older students emphasize cognitive prompting and metacognitive reflection. These differences underscore how instructional dialogue dynamically reorganizes according to learners’ developmental and curricular contexts, and how network models can expose the implicit pedagogical logic underlying adaptive tutoring. In a related attempt to visualize how instructional strategies shape learning processes, \cite{Nguyen2024} uses ordered co-occurrence networks of practices to analyze how reflective design activities and self-regulated learning strategies are temporally interconnected during AI-supported design thinking.

The professional learning perspective is further advanced in \cite{Li2024_b}, which uses ENA within a quantitative ethnography framework to examine pre-service teachers’ competencies and challenges while designing inquiry-based integrated STEM lessons. Nodes represent categories such as “pedagogical content knowledge (PCK) limitation,” “design difficulty,” and “curricular connection,” while edges reflect their co-occurrence in reflective narratives. The resulting networks highlight persistent challenges in the Explore, Elaborate, and Evaluate stages of the 5E instructional model commonly used in STEM lesson design, revealing that design obstacles often cluster around weak PCK integration. By mapping these patterns, the study identifies actionable leverage points for targeted teacher education interventions.

\begin{figure}[ht]
    \centering
    \includegraphics[width=0.5\textwidth]{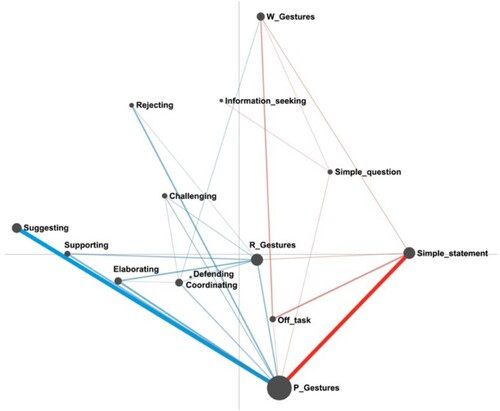}
    \caption{Network extracted from \cite{Rachmatullah2024}: Epistemic network constructed from coded discourse and gestures during collaborative science problem solving. Nodes represent communicative and cognitive practices, and edges indicate their co-occurrence, revealing the structure of students’ multimodal reasoning processes.}
    \label{example_abi_pract_net}
\end{figure}

A complementary angle is offered by \cite{Rachmatullah2024}, who analyze students’ multimodal practices—speech and gestures—during collaborative problem-solving in science. Using ENA with correlation-based weighting (Phi coefficients), they construct networks linking categories of verbal and gestural behavior. While the analysis remains descriptive, it demonstrates the potential of network visualization to capture the coordination of embodied and discursive modes of participation, showing, for instance, how explanatory gestures often co-occur with elaborative speech in high-performing groups. Even without deep network metrics, this study exemplifies how ENA can bridge multimodal learning analytics and interactional research in science education. An example of these networks is shown in Figure \ref{example_abi_pract_net}.

Taken together, these works situate practice-based learning as a networked phenomenon, where meaningful participation arises from recurrent co-patterns of communicative, cognitive, and affective actions. Across classroom, online, and professional contexts, [AN-P] studies reveal that competence is not a linear trait but an emergent property of connected practices. Methodologically, this corpus also reflects a trend toward hybrid analytics—integrating qualitative ethnography, machine learning, and ENA—to capture the temporal and relational complexity of science learning in action. In doing so, the [AN-P] literature extends the reach of network analysis beyond conceptual knowledge into the embodied, situated, and interpersonal dimensions of scientific practice.

\section{Tools from complex networks used in science education research so far}

\begin{figure}[ht]
    \centering
    \includegraphics[width=0.5\textwidth]{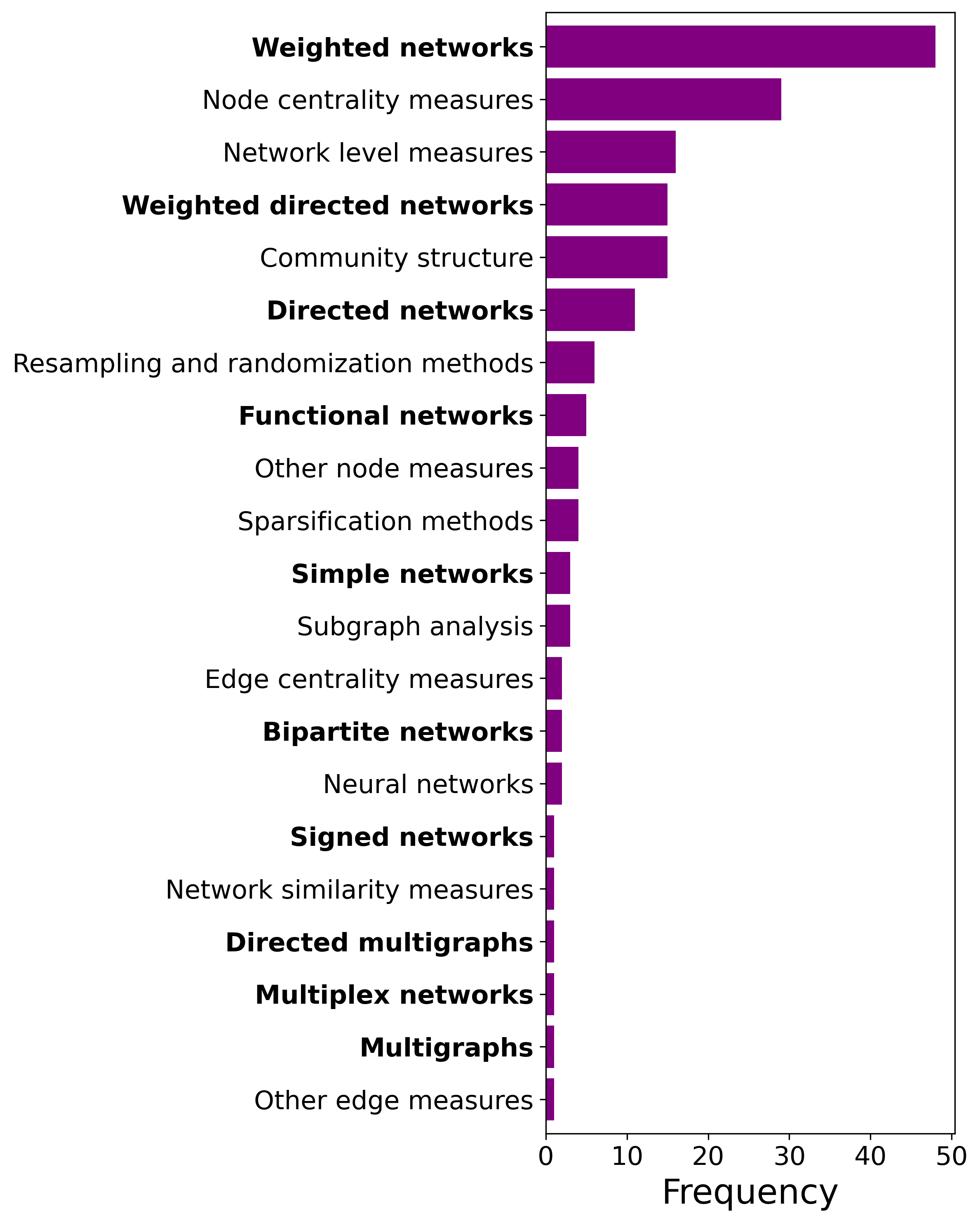}
    \caption{Frequency of network tools used in the articles covered in this review. Bold labels refer to types of networks used.}
    \label{fig:network_tools}
\end{figure}

The frequency of network tools used in the corpus of articles reviewed here can be seen in Fig.~\ref{fig:network_tools}. 

\paragraph{Types of network representations}

Across the corpus, there is a clear dominance of weighted network representations (48 papers). This shows that most studies consider interaction strength as essential information rather than relying on binary structure. A substantial subset additionally incorporates directionality, either in weighted directed networks (15) or directed networks without weights (11), reflecting the importance of asymmetric relations in many educational and behavioral settings. Only a few studies use simple (unweighted, undirected) networks (3), indicating that purely topological abstractions are now comparatively uncommon.

A smaller but meaningful group employs more specialized representations: functional networks (5), where links reflect statistical similarity rather than explicit interactions; bipartite networks (2), often used to represent two-mode data; and rare cases of signed networks (1), multiplex networks (1), multigraphs (1), or directed multigraphs (1). These more complex frameworks appear only sporadically, suggesting that multi-layer, antagonistic, or multi-edge structures remain underexplored in this literature.

\paragraph{Network tools and analytical methods}

Regarding analytical techniques, the most common tools are node centrality measures (29) and community-structure detection methods (15), showing a strong emphasis on identifying influential actors and uncovering group-level organization. Network-level summary measures (16) are also widely used, serving as a baseline characterization of global topology.

Intermediate-frequency methods include resampling and randomization procedures (6) for statistical validation, sparsification methods (4) that reduce network density while preserving structure, and additional node-level metrics beyond standard centralities (4). Less frequent are subgraph or motif analyses (3), edge-level centrality (2), other edge metrics (1), and network similarity measures (1). These edge-based and comparative approaches remain uncommon relative to node-based metrics.

Overall, the literature is methodologically concentrated: most studies rely on weighted (and often directed) representations combined with classic node centralities and community detection, while more structurally rich network types and edge- or motif-focused methods are used only rarely.

\section{Coauthorship and citation network}

\paragraph{Coauthorship network}

The coauthorship network contains 204 authors connected by 404 coauthorship edges. The network is fragmented into 52 connected components, most of which correspond to a single category of research articles. Only five components include authors publishing in two different categories. The largest component contains 30 authors and includes papers in the SN and CN categories. It is followed by a component of 20 authors (CN and BN) and another with 11 authors dedicated exclusively to SN. The remaining components contain between one and seven authors, with the majority consisting of pairs of authors (see the data and code in the repository associated with this paper~\cite{fernandezgracia2026cnser}).

These results indicate that researchers applying complex networks to science education do not form a highly cohesive collaboration community. Instead, the field appears fragmented into many small and largely disconnected groups, with limited collaboration across topical categories.

\paragraph{Citation network}

The citation network contains a largest connected component comprising 6,592 authors linked by more than 32,000 citation ties. Community detection within this component reveals 21 communities, with three major clusters structured around relatively distinct intellectual traditions. One cluster is centred on methodological and theoretical contributions to network science (e.g., Newman, Wasserman, Faust), another around learning sciences and knowledge integration (e.g., Chi, Krajcik), and a third around argumentation and epistemology in science education (e.g., Shaffer, Erduran).

The in-degree distribution of the network is highly skewed: half of the authors receive three or fewer citations within the corpus, whereas only a small number of highly cited authors reach around 60–70 citations. In addition, the analysis of centrality measures reveals a small set of authors with elevated betweenness centrality that connect otherwise weakly linked parts of the citation network.

These patterns indicate that the intellectual landscape of research on complex networks in science education is organised around several relatively distinct traditions that are linked through a limited number of bridging figures. When considered together with the results from the coauthorship network, the citation network provides complementary information: while the collaboration network reveals a field fragmented into many small and largely disconnected research groups, the citation network shows that these groups nevertheless draw on a partially shared intellectual foundation. Overall, the combined picture suggests that the field is still in a phase of consolidation, with multiple sub-communities that employ similar network-analytic approaches but remain only loosely integrated in terms of collaboration and citation structure.

\section{Potential educational applications of complex networks yet to be identified}

A recurring pattern across the literature is that many studies already collect or report data whose structure naturally invites more advanced network-science approaches, yet the analyses remain within static or single-layer frameworks. Numerous papers construct networks at multiple points in time, distinguish different contexts for the same edges (e.g., in-class vs. out-of-class interactions), or analyze different cohorts, tasks, or subnetworks. Several works compare networks across courses or years, examine parallel networks for men and women, or describe networks with colored edges that are essentially multiplex layers. Others include node-level covariates, functional relations (e.g., fMRI-based networks), or sets of networks constructed from related tasks. However, with few exceptions, these datasets are not analyzed with the temporal, multilayer, or coevolutionary tools that would fully exploit their structure.

This predominance of static and descriptive analyses should not be interpreted solely as a lack of methodological sophistication. It also reflects the specific nature of many phenomena studied in science education, where learning is not primarily driven by fast collective dynamics or direct diffusion processes. Knowledge and understanding are not transmitted between learners in the same way that opinions, behaviors, or diseases spread in classical contagion models. Instead, learning typically unfolds as an intraindividual process, strongly shaped by instruction, assessment, and context, and only indirectly mediated by social interaction. From this perspective, the value of network approaches lies in capturing processes of structural reorganization over time, such as the gradual restructuring of conceptual networks or the stabilization of participation patterns in learning communities, and in revealing how changes in one part of the educational system propagate, constrain, or reshape structures in other parts of the system.

These observations highlight substantial opportunities for methodological development. Temporal-network models would allow researchers to move beyond retrospective comparisons of static snapshots, capturing instead the ordering, duration, and evolution of interactions. The frequent appearance of multiple time points—often treated as independent networks—suggests that methods for temporal motifs, persistence, burstiness, and link turnover could yield new insights into learning processes, community formation, or the development of conceptual understanding.

Similarly, many studies implicitly contain multiple layers of interaction (different tasks, sections, social contexts, or demographic groups). Formal multilayer and multiplex network frameworks would enable quantifying inter-layer dependencies, cross-context coupling, and the alignment or misalignment of students’ roles across tasks. Even simple colored-edge data could be reframed as a multiplex network, allowing researchers to investigate layer dominance, redundancy, or complementarities in learning activities. Where several groups perform the same task or where subnetworks are defined for different phases of an activity, multilayer models could characterize how shared structure emerges, how collaboration strategies propagate, or how group-level differences arise. In this sense, multilayer modeling is particularly well suited to science education research, as it allows comparing how the same individuals, concepts, or practices are structurally positioned across instructional contexts, pedagogical designs, or assessment conditions, rather than assuming a single homogeneous learning environment.

Another untapped opportunity lies in mechanistic or generative models on networks. Many datasets hint at dynamical processes—for example, changes across tasks, repeated measurements across semesters, or functional networks from brain activity—but the analyses remain purely descriptive. Incorporating models of diffusion, opinion dynamics, threshold-based adoption, or reinforcement processes could help explain how conceptual change spreads, how collaboration patterns stabilize, or how learning strategies propagate through a group. Likewise, coevolutionary models, where node states (e.g., knowledge, engagement) and network ties influence each other, would match the natural feedbacks in learning environments far better than static summaries.

Finally, several papers rely on bibliometric or software-generated networks using tools such as VOSviewer, Citespace, or ENA. These approaches, while valuable, often treat networks as visualization devices rather than analytical objects. Adopting richer network measures—temporal stability, layer-aligned centralities, signed or weighted dynamics, higher-order interactions, or network similarity metrics—could deepen the interpretive and predictive power of such studies. Beyond individual analyses, network and complex-systems approaches also offer a way to bring coherence to a field characterized by strong contextual dependence, limited standardization, and, sometimes, highly fragmented instructional practices. By mapping connections across studies, curricula, concepts, and educational designs, network-based analyses can help identify common structural patterns, recurrent mechanisms, and organizing principles that are otherwise obscured by the diversity of local contexts and isolated interventions.

In summary, much of the articles in the field already gather information that naturally supports temporal, multilayer, coevolutionary, higher-order, and mechanistic analyses, yet these tools are rarely applied. At the same time, the potential of network approaches in science education lies not in forcing dynamic models where they are theoretically unwarranted, but in leveraging networks as integrative frameworks that connect cognitive, social, and institutional dimensions of learning. This gap represents a major frontier for science education research: moving from static, single-layer descriptions toward dynamic, structurally rich, theoretically grounded network models, that better reflect the complexity of real learning processes.

\section{Conclusions}

This review provides the first systematic and integrative synthesis of how complex network approaches have been used in science education research across conceptual, social, epistemic, and bibliometric domains. By analysing 82 publications from nine leading journals, we identify four broad families of network applications—concept networks, social networks, bibliographic networks, and abilities/practices networks—each addressing different facets of teaching, learning, and knowledge production in science education.

Across these categories, several patterns emerge. First, most studies employ networks primarily as descriptive or exploratory tools, using classical graph-theoretical measures (centralities, communities, density) to reveal structural patterns in conceptual organisation, social interactions, and the structure of the research field. Concept networks, in particular, dominate the literature, reflecting the centrality of understanding how learners connect ideas and how these connections evolve. Social network research shows strong and consistent links between students’ positions in interaction networks and their engagement, achievement, and persistence in STEM pathways. Bibliographic networks provide a meta-perspective on how the field itself is organised, while studies of reasoning and practice illustrate the growing use of ENA to quantify epistemic and behavioural structures in learning.

Second, despite the methodological diversity of the corpus, many studies implicitly contain richer structural information—temporal sequences, multiple contexts, multimodal data, or parallel groups—that is not fully leveraged. Temporal, multilayer, and coevolutionary network models remain almost entirely absent, even though the data reported in many papers naturally support them. Similarly, mechanistic or generative models on networks, which could provide explanatory leverage on learning dynamics, collaboration patterns, or conceptual change, are rarely used. This indicates a substantial methodological gap between the potential of network science and its current application in science education.

Third, while ENA has become a prominent tool in recent years, its use is often limited to visualising co-occurrence structures rather than advancing theory. The field would benefit from clearer distinctions between epistemic, semantic, conceptual, and interaction networks, as well as from stronger theoretical integration connecting network patterns to models of learning, reasoning, or social participation.

Our analysis of the coauthorship and citation networks of the reviewed corpus further clarifies how the field is organised as a research community. The coauthorship network reveals a highly fragmented collaboration structure, composed of many small and largely disconnected groups that tend to work within specific topical categories. In contrast, the citation network shows that these groups are embedded in a broader intellectual landscape structured around several major traditions, including methodological work in network science, research in the learning sciences, and studies of argumentation and epistemology in science education. While these traditions remain partially distinct, they are connected through a small set of bridging authors and shared methodological references. Taken together, these results suggest that the field is intellectually connected through common conceptual and methodological foundations, but socially dispersed in terms of direct collaboration.

Overall, our review reveals a growing but uneven landscape. Network approaches have contributed important insights into conceptual development, classroom interactions, and the organisation of science education research itself. Yet moving forward, the field has the opportunity to adopt more structurally sophisticated, dynamic, and theory-driven network models. We argue that embracing multilayer, temporal, generative, and higher-order frameworks, together with clearer theoretical grounding, can transform network analysis from a descriptive methodology into a powerful engine for explaining and predicting learning phenomena in science education.

\section{Code and data availability statement}

All data and code needed for this review can be accessed in the repository in reference~\cite{fernandezgracia2026cnser}.

\section{Acknowledgements}

JFG acknowledges partial support from project CSxAI (PID2024-157526NB-I00) funded by MICIU/AEI/10.13039/501100011033/FEDER, UE.

% \bibliography{biblio}


\end{document}